\documentclass[aps,prd,twocolumn,
superscriptaddress,floatfix,showpacs,showkeys,preprintnumbers,
  amsmath,amssymb,
  aps,
  prd
]{revtex4-2}

\usepackage{graphicx}
\usepackage{dcolumn}
\usepackage{bm}
\usepackage{hyperref}

\usepackage[english]{babel}
\usepackage{xcolor}
\usepackage{mathtools}
\usepackage{pdfsync}
\usepackage{amssymb}
\usepackage{amsmath}

\usepackage{slashed}
\usepackage[active]{srcltx}
\RequirePackage{mathptmx}      

\newcommand{\Li}{\operatorname{Li}}
\def\d{\hbox{{d}\kern-.20em\hbox{l}}}

\def \matrix #1 {\left(\begin{array}{cc} #1 \end{array}\right)}

\def\II{\hbox{{1}\kern-.25em\hbox{l}}}

\newcommand{\MS}{\overline{\mathrm{MS}}}
\newcommand{\p}{\mathbf p}

\begin{document}


\title{
Three-loop off-forward evolution kernel for axial-vector operators in Larin's scheme}


\author{Vladimir M. Braun}
   \email{vladimir.braun@ur.de}
   \affiliation{Institut f\"ur Theoretische Physik, Universit\"at
   Regensburg, D-93040 Regensburg, Germany}

\author{Alexander N. Manashov}
   \email{alexander.manashov@desy.de}
\affiliation{II. Institut f\"ur Theoretische Physik, Universit\"at Hamburg
   D-22761 Hamburg, Germany}
\affiliation{
   St.Petersburg Department of Steklov
Mathematical Institute, 191023 St.Petersburg, Russia}

\author{Sven-O. Moch}
\email{sven-olaf.moch@desy.de} \affiliation{II. Institut f\"ur Theoretische Physik, Universit\"at Hamburg
   D-22761 Hamburg, Germany}

 \author{Matthias Strohmaier}
   \email{strohmaiermatthias@gmail.com}
\affiliation{Institut f\"ur Theoretische Physik, Universit\"at
   Regensburg, D-93040 Regensburg, Germany}

\begin{abstract}
Evolution equations for leading twist operators in high orders of perturbation theory
can be restored from the spectrum of anomalous dimensions and the calculation of the special conformal
anomaly at one order less using conformal symmetry of QCD at the Wilson-Fisher critical
point at non-integer $d=4-2\epsilon$ space-time dimensions.
In this work we generalize this technique to axial-vector operators.
We calculate the corresponding three-loop evolution kernels in Larin's scheme and
derive explicit expressions for the finite renormalization kernel that describes
the difference to the vector case to restore the conventional $\MS$-scheme.
The results are directly applicable to deeply-virtual Compton scattering and
the transition form factor $\gamma^*\gamma\to\pi$.
\end{abstract}



\preprint{ \textmd{\normalsize DESY--20--236}}


\maketitle

%
%

\section{Introduction}\label{sec:intro}

The QCD description of hard exclusive reactions in the framework of collinear factorization involves
matrix elements of leading-twist operators between hadron states with different momenta ---
generalized parton distributions (GPDs) or light-cone distribution amplitudes (LCDAs).
Such processes are attracting increasing attention because they provide complementary information on
the hadron structure as compared to inclusive reactions, and because of the very high quality of experimental
data from the JLab 12 GeV upgrade~\cite{Dudek:2012vr}, SuperKEKB
\cite{Kou:2018nap}, and, in the future, from the EIC~\cite{Accardi:2012qut}.
The main motivation for this study is provided by the applications to deeply-virtual Compton scattering (DVCS),
but the results are also relevant for reactions of the type $\gamma\gamma^\ast \to \pi$ etc.

The theoretical description of such reactions has to match the experimental accuracy. In particular evolution equations for GPDs and
related quantities have to be derived to similar precision as for the usual parton distributions, currently completely known at NNLO
(three-loop)~\cite{Moch:2004pa,Vogt:2004mw}. The difference in these two cases is that for GPDs (and LCDAs) mixing with operators
containing total derivatives must be taken into account. The complete NLO (two-loop) evolution kernels for GPDs have been calculated long
ago~\cite{Mueller:1997ak,Belitsky:1998vj,Belitsky:1998gc} using an approach developed by D.~M\"uller~\cite{Mueller:1991gd}. These results
were later rederived and confirmed~\cite{Braun:2014vba,Braun:2019qtp} by a somewhat different technique~\cite{Braun:2013tva,Braun:2014vba}
that makes use of (exact) conformal symmetry of QCD at the Wilson-Fisher critical point in non-integer $d=4-2\epsilon$ dimensions. Since
renormalization constants in dimensional regularization with minimal subtraction, by construction, do not depend on the space-time
dimension, evolution equations in physical QCD at $d=4$ are the same as in the critical theory at $d=4-2\epsilon$ and possess a ``hidden''
conformal symmetry: The evolution kernels in QCD commute with the generators of conformal transformations. These generators in the
interacting theory are modified (``deformed'') by quantum corrections and the corresponding modification can be calculated order by order
in perturbation theory using conformal Ward identities~\cite{Belitsky:1998gc,Braun:2013tva,Braun:2014vba,Braun:2016qlg}. With this
approach, the three-loop evolution kernels for GPDs~\cite{Braun:2017cih} and the corresponding two-loop coefficient
functions~\cite{Braun:2020yib} for DVCS were calculated for flavor-nonsinglet vector-like distributions.

The extension of this technique to axial-vector distributions requires special considerations due to known issues with the definition of
the $\gamma_5$-matrix in non-integer dimensions. Having in mind applications to two-photon reactions the well-known prescription by
Larin~\cite{Larin:1993tq} suggests itself, see also
Refs.~\cite{Akyeampong:1973xi,Larin:1991tj,Zijlstra:1992kj,Matiounine:1998re,Moch:2015usa}. In this work we study axial-vector operators
defined in Larin's scheme under conformal transformations at NLO and, following the method developed in~\cite{Braun:2017cih}, use this
result to restore the corresponding three-loop evolution kernel. In QCD it is natural to fix the conventional $\MS$ renormalization scheme
such that the evolution equations for vector and axial-vector flavor-nonsinglet operators coincide identically. Starting from Larin's
prescription this requires a finite renormalization which is then used to redefine the coefficient functions. This finite renormalization
kernel is known for the forward case to three loops~\cite{Moch:2014sna}. In this paper we derive the corresponding expression for the
general off-forward case.

The presentation is organized as follows. In Sect.~\ref{sec:def} we introduce axial-vector light-ray operators of leading twist in
non-integer dimensions. Sect.~\ref{sect:symandren} contains a brief discussion and comparison of the symmetries and renormalization
properties of vector and axial-vector operators.  We present our results for the two-loop conformal anomaly for axial-vector operators in
Sect.~\ref{sec:two-loop-anomaly} and for the three-loop evolution kernel in Sect.~\ref{sect:threeH}. The rotation matrix from Larin's
scheme to the $\MS$-scheme is given in Sect.~\ref{sect:matching}. We conclude in Sect.~\ref{sect:summary} and the rotation matrix in the
local operator basis is presented in App.~\ref{sect:appendix}.

\section{Definition}\label{sec:def}

GPDs (and LCDAs) are defined as off-forward matrix elements of leading-twist light-ray operators. For the vector case
\begin{align}
{\mathcal O}_V(z_1,z_2) &=\bar q(z_1n) \gamma_{+} [z_1n,z_2n] q(z_2),
\label{vector}
\end{align}
where $n^\mu$ is an auxiliary light-like vector, $n^2 = 0$, $z_{1,2}$ are real numbers, $\gamma_+ = \slashed{n} = n^\mu\gamma_\mu$
and $[z_1n,z_2n]$ is the Wilson line.
The light-ray operator~\eqref{vector} can be viewed as the generating function for local operators,
${\mathcal O}_{V}^{\mu_1\dots\mu_N}$ that are symmetric and traceless in all indices $\mu_1\dots\mu_N$.

The corresponding axial-vector light-ray operator in four dimensions is naturally given by
\begin{align}
{\mathcal O}_{5}(z_1,z_2) &=\bar q(z_1n)\gamma_+\gamma_5 [z_1n,z_2 n]q(z_2n).
\label{O5}
\end{align}
However, this definition is not suitable for theories in non-integer $d=4-2\epsilon$ dimensions as the $\gamma_5$-matrix is not defined.

Note that in applications to two-photon reactions one needs the operator product expansion (OPE) of
two vector (electromagnetic) currents, which is perfectly well defined in non-integer dimensions. The
products of $\gamma$-matrices which occur in loop diagrams can be reduced to the basis of antisymmetric products
\begin{align}
\Gamma_{\mu_1\dots\mu_n}= \gamma_{[\mu_1}\dots\gamma_{\mu_n]},
\end{align}
where $[\ldots]$ stands for antisymmetrization in the enclosed Lorentz indices. It is easy to see that for leading twist operators the
antisymmetric products of more than three $\gamma$-matrices cannot appear (in other words, there are no evanescent operators) so that only
one light-ray operator can contribute in addition to \eqref{vector}:
\begin{align}\label{def:OmnLR}
{\mathcal O}_{\mu\nu}(z_1,z_2)=\bar q(z_1n)\Gamma_{\mu\nu\alpha}n^\alpha [z_1n,z_2 n]q(z_2n).
\end{align}
In four dimensions this operator can be rewritten in terms of the axial-vector operator \eqref{O5}
\begin{align}
{\mathcal O}_{\mu\nu}(z_1,z_2) &= i\epsilon_{\mu\nu\alpha\beta} n^\alpha\frac\partial{\partial {n_\beta}}
\int_0^1 du\, {\mathcal O}_{5}(z_1 u,z_2 u)+\dots.
\end{align}
where ellipses stand for higher-twist terms.

The addenda appear because the operator \eqref{def:OmnLR} does not have a definite twist yet.
The reason is that when going over to local operators,
\begin{align}
   n^\alpha n^{\nu_1}\ldots n^{\nu_{N-1}}\bar q(0) \Gamma_{\mu\nu\alpha}D_{\nu_1} \ldots D_{\nu_{N-1}} q(0),
\end{align}
the multiplication with the auxiliary light-like vector $n^\mu$ does not yield a traceless result
in pairs of indices $(\mu, \nu_k)$ and $(\nu,\nu_k)$. An additional subtraction is needed to separate the
leading twist part. The corresponding condition in the light-ray operator formalism is that
the leading twist-two part of the operator ${\mathcal O}_{\mu\nu}(z_1,z_2)$ must obey the constraint
\begin{align}\label{constraint}
\partial^\mu\, {\mathcal O}_{\mu\nu}(z_1,z_2)\big|_{l.t.} =0.
\end{align}
%
For a generic matrix element between states with different momenta this constraint
reduces to
\begin{align}
\Delta^\mu\langle P'| {\mathcal O}_{\mu\nu}(z_1,z_2) |P\rangle=0\,,\qquad \Delta^\mu = (P'-P)^\mu.
\end{align}
Since in addition (by construction)
\begin{align}
 n^\mu\langle P'| {\mathcal O}_{\mu\nu}(z_1,z_2) |P\rangle=0,
\end{align}
it follows that the leading-twist part of the operator ${\mathcal O}_{\mu\nu}(z_1,z_2)$
corresponds to the transverse components with respect to the $n, \Delta$ plane.
Let
\begin{align}
\Delta^\mu = \alpha n^\mu+ \beta \bar n^\mu,
\end{align}
with $\bar n^2= n^2=0$, $(n\bar n)=1$,
and choose two orthogonal unit vectors in transverse directions, $a^\mu$ and $b^\mu$, such that
$(a\cdot n)= (a \cdot\bar n)=0$ and $(b\cdot n)= (b \cdot\bar n)=0$.

The leading-twist-two axial-vector operator in non-integer dimensions can
be defined as
\begin{align}\label{def:OA}
{\mathcal O}_A(z_1,z_2) & =a^\mu b^\nu {\mathcal O}_{\mu\nu}(z_1,z_2)
=\bar q(z_1n) \Gamma_+ [z_1n,z_2n] q(z_2),
\end{align}
where
\begin{align}
\Gamma_+ =  a^\mu b^\nu \Gamma_{\mu\nu\alpha} n^\alpha.
\label{Gamma}
\end{align}
Note that in four dimensions
 $ a^\mu b^\nu \epsilon_{\mu\nu\alpha\sigma} n^\alpha \sim n_\sigma$
so that $\Gamma_+ \sim i\gamma_+\gamma_5$
and the operator in \eqref{def:OA} reduces to the one in \eqref{O5}, as desired.
Our definition is, of course, a version of Larin's scheme~\cite{Larin:1993tq}.

\section{Renormalization and symmetries}\label{sect:symandren}

Here and in what follows we consider QCD in $d=4-2\epsilon$ dimensions and tacitly
imply that all operators are renormalized in the $\MS$-scheme.

Renormalized  light-ray operators satisfy the renormalization group (RG) equations
\begin{align}\label{VARGE}
\Big(\mu\partial_\mu +\beta(a)\partial_{a}+ \mathbb{H}_{\p}(a)\Big) {\mathcal O}_{\p}(z_1,z_2)=0,
\end{align}
where $ \p = V,\,A,$ the strong coupling is $a=\alpha_s/(4\pi)$ and $\beta(a)$ is the $d$-dimensional beta function,
\begin{align}
\beta(a)= -2a\big(\epsilon+\beta_0 a +\beta_1 a^2+O(a^3)\big),
\end{align}
$\beta_0=11/3N_c-2/3 n_f$, etc. for an $\text{SU}(N_c)$ gauge theory with $n_f$ quark flavors.
At the critical point the  $d$-dimensional beta function vanishes. This can be achieved either by
fine-tuning the coupling $a \mapsto a_\ast$ for fixed $\epsilon = (4-d)/2$, or fine-tuning $\epsilon \mapsto \epsilon_\ast = -\beta_0 a^2 -
\beta_1 a^3 -\ldots$ for a fixed value of the coupling. In what follows we use both notations intermittently.

The evolution kernels $\mathbb{H}_{\p}(a)$ are integral operators in $z_1,z_2$,
\begin{align}
\mathbb{H}_{\p}(a)=\sum_{\ell=1}^\infty a^\ell\, \mathbb H^{(\ell)}_{\p}.
\end{align}
They can be written in the form
\begin{align}
[\mathbb H^{(\ell)}_{\p} f](\boldsymbol z) = \int_0^1 d\alpha d\beta\, h_{\p}^{\ell}(\alpha,\beta)\,f(z_{12}^\alpha,z_{21}^\beta),
\end{align}
where $\boldsymbol z$ abbreviates the set of $z_{1,2}$, i.e., $\boldsymbol z \equiv\{z_1,z_2\}$ and
\begin{align}
z_{12}^\alpha\equiv z_1\bar \alpha +z_2\alpha, &&\bar\alpha\equiv 1-\alpha.
\end{align}
The one-loop  kernels for the vector and axial-vector operators coincide, $\mathbb H^{(1)}_{V}= \mathbb H^{(1)}_{A} $, and are given
by the following expression \cite{Balitsky:1987bk}:
\begin{multline}\label{Honeloop}
{}[\mathbb{H}_{\p}^{(1)}\!f](\boldsymbol z)  =4C_F\biggl \{\frac12 f(\boldsymbol z)
-\int_0^1\!d\alpha\int_0^{\bar\alpha}\!d\beta  f(z_{12}^\alpha,z_{21}^\beta)
\\
+\int_0^1d\alpha\frac{\bar\alpha}{\alpha}\Big(2f(\boldsymbol z)-f(z_{12}^\alpha,z_2)-f(z_1,z_{21}^\alpha)\Big) \biggr \}\,.
\end{multline}
At the classical level, vector \eqref{vector} and axial-vector light-ray operators \eqref{def:OA} transform  under
conformal transformations in the same way. As a consequence, the one-loop evolution kernels in both cases
commute with the canonical generators of the collinear conformal subgroup,
\begin{align}
[S_{\pm,0},\mathbb H_{\p}^{(1)}]=0,
\label{commute1}
\end{align}
where
\begin{align}
S_-&=-\partial_{z_1}-\partial_{z_2},
\notag\\
S_0\, &= z_1\partial_{z_1}+ z_2\partial_{z_2}+2,
\notag\\
S_+ &= z_1^2\partial_{z_1}+ z_2^2\partial_{z_2}+2z_1+2z_2.
\label{canon}
\end{align}
This property follows from the conformal invariance of the  QCD Lagrangian at the classical level. Beyond tree level the scale and
conformal symmetries are  broken by quantum corrections. In  non-integer dimensions, however, there exists a nontrivial fixed point,
$a=a_*$, such that $\beta(a_*)=0$, so that for this special choice of the coupling both scale and conformal invariance of the theory  are
restored. The symmetry generators  in the critical theory, $\mathrm S_{\pm,0}$ , satisfy the usual $\text{SL}(2)$ algebra but differ from
the canonical generators~\eqref{canon} by quantum corrections.
These ``deformed'' symmetry generators commute with the evolution kernels
\begin{align}\label{SHcomm}
[\mathrm S^{\p}_\alpha (a_*),\mathbb H_{\p}(a_*)]=0.
\end{align}
 It can be shown, see Ref.~\cite{Braun:2016qlg} for details, that this modification affects
 the generators of dilatations and special conformal transformations which take the form
\begin{align}\label{Sfull}
\mathrm S^{\p}_0\, &= S_0 +\left(-\epsilon+\frac12 \mathbb H_{\p}(a_*)\right),
\notag\\
\mathrm S^{\p}_+\, &= S_+ +(z_1+z_2)\left(-\epsilon+\frac12 \mathbb H_{\p}(a_*)\right) +z_{12}\Delta_{\p}(a_*),
\end{align}
whereas the generator $S_-$ (corresponding to translations along the light-ray)
does not receive any corrections, $\mathrm S^{\p}_- = S_- $.
These expressions are valid for both cases, $\p = V,A$.
Note that the modification of the generator of dilatations is expressed in terms of the evolution kernel
whereas for the conformal transformations there is an additional contribution.
This additional term is usually referred to as the conformal anomaly.
It can be calculated order-by-order in perturbation theory,
\begin{align}
\Delta_{\p}(a_*)=\sum_\ell a_*^\ell \Delta_{\p}^{(\ell)},
\end{align}
from the conformal Ward identities for the corresponding light-ray operators,
see~\cite{Belitsky:1998gc,Braun:2016qlg} for details. At one-loop order, the conformal anomaly for the vector and axial-vector operators
coincide, $\Delta^{(1)}_V=\Delta^{(1)}_A$. The corresponding expression was first obtained in Ref.~\cite{Belitsky:1998gc}:
\begin{align}
[\Delta^{(1)}_{\p} f](\boldsymbol z) &=
\int_0^1d\!\alpha\, \omega^{(1)}(\alpha)
 [f(z_{12}^\alpha,z_2)-f(z_1,z_{21}^\alpha)],
\end{align}
where the weight function $\omega^{(1)}$ reads
\begin{align}
\omega^{(1)}(\alpha)=-2C_F\left(\frac{\bar\alpha}\alpha+\ln\alpha\right).
\end{align}
The two-loop conformal anomaly for the vector case, $\Delta^{(2)}_V$, was derived in~\cite{Braun:2016qlg},
the resulting expression being too lengthy to be presented here.

It is convenient to write the evolution kernels, $\mathbb H_{\p}$, as a sum of two terms of which the first one is invariant
and the second one non-invariant with respect to the canonical symmetry transformations. Suppressing the $V/A$ subscript,
we define
\begin{align}
\label{Hseparation}
\mathbb H =\mathbb H^{\text{inv}} + \mathbb H^{\text{non-inv}}, && [S_\alpha,\mathbb H^{\text{inv}}]=0.
\end{align}
It was shown~\cite{Mueller:1991gd,Belitsky:1998gc,Braun:2014vba} that the operator $ \mathbb H^{\text{non-inv}}$ at
$\ell$-loop order is completely determined (up to invariant terms) by the conformal anomaly at one order less, $\ell-1$ .
Once this non-invariant operator is fixed, the invariant part $\mathbb H^{\text{inv}}$ can
be restored from the corresponding anomalous dimensions (at $\ell$ loops). These are  known
for vector and axial-vector operators
(in Larin's scheme) to three-loop accuracy~\cite{Moch:2004pa,Moch:2014sna}.

Since  $\mathbb H^{(1)}_{V}= \mathbb H^{(1)}_{A} $ and $\Delta_V^{(1)}=\Delta_A^{(1)}$, the
generators of special conformal transformations for vector and axial-vector operators coincide to this accuracy. From the
commutation relations \eqref{commute1} it follows then that the leading two-loop term for the difference of the evolution kernels,
\begin{align}\label{HAVdiff}
\mathbb H_{A-V}(a) & = \mathbb H_A(a)-\mathbb H_V(a) = a^2\mathbb H_{A-V}^{(2)} +O(a^3),
\end{align}
is a canonically invariant operator. It is completely determined by its spectrum which is given by the difference
of anomalous dimensions of vector and axial-vector operators, $\gamma_V(N)-\gamma_A(N)$, where $N$ denotes the spin of the operator.
We obtain after a short calculation
\begin{align}\label{deltaHAV}
[\mathbb H_{A-V}^{(2)} f](\boldsymbol z)= 16C_F \beta_0\int_0^1d\alpha\int_0^{\bar\alpha}d\beta f(z_{12}^\alpha,z_{21}^\beta).
\end{align}
In order to calculate the difference of the vector an axial-vector kernels to $O(a^3)$ one needs
the difference in the corresponding conformal anomalies at two loops.

\section{Conformal anomaly for axial-vector operators to $O(a^2)$}
\label{sec:two-loop-anomaly}

The calculation of the conformal anomaly for vector operators is discussed at length in \cite{Braun:2014vba,Braun:2016qlg}.
The only modification for the axial-vector operators is to replace the
$\gamma_+$ matrix by $\Gamma_+$ defined in Eq.~\eqref{Gamma} in the operator vertex
in the corresponding diagrams. Simplifying the numerators, one uses the following properties of the $\gamma_+$ matrix
in $d=4-2\epsilon$
\begin{align}
\gamma_+\gamma_+=0, && \gamma^\mu \gamma_+\gamma_\mu=-2(1-\epsilon)\gamma_+\,.
\end{align}
The corresponding identities for $\Gamma_+$  take the form
\begin{align}
\gamma_+\Gamma_{+} =\Gamma_{+}\gamma_+=0,
&&
 \gamma^\mu \Gamma_{+}\gamma_\mu = 2(1+\epsilon)\Gamma_{+}\,.
\end{align}
The diagrams to be calculated are shown in Fig.~2,
Ref.~\cite{Braun:2016qlg}. One is interested in the residues of the simple poles in $\epsilon$.
It is easy to see that the replacement $\gamma_+\to\Gamma_{+}$
does not affect the diagrams with one interacting quark in Fig.~2(a-g). For the  remaining diagrams in
Fig.~2(h-g)  the modifications due to the substitution  $\gamma_+\to\Gamma_{+}$ can easily be tracked down
and are related to the factors $(1-\epsilon)^k$ vs.  $(1+\epsilon)^k$ arising in the calculation.
The expression for a generic diagram has the form
\begin{align*}
\gamma^{\alpha_1}\ldots\gamma^{\alpha_{2k}} \Gamma \gamma_{\beta_1}\ldots\gamma_{\beta_{2m}}
\left(\frac1{\epsilon^2} (T_1)^{\beta_1\ldots\beta_{2m}}_{\alpha_1\ldots\alpha_{2k}}+
\frac1{\epsilon} (T_2)^{\beta_1\ldots\beta_{2m}}_{\alpha_1\ldots\alpha_{2k}}\right),
\end{align*}
where $\Gamma=\gamma_+\text{ or } \Gamma_{+}$ and $T_{1,2}$ are certain tensors which depend on the external momenta.
Since we are interested in the $1/\epsilon$ pole only,
contracting the Lorentz indices on the string of $\gamma$-matrices with $T_2$ one can use the four-dimensional algebra and replace
$\Gamma_{+}$ by $i\gamma_+\gamma_5$. The $1/\epsilon$ contributions due to $T_2$  are therefore the same for the
vector and axial-vector operators. Hence one  needs to consider the double-pole contributions of $T_1$ only,
which are related to divergent subgraphs and are easy to calculate. Such double-pole contributions for a given diagram take the form
\begin{align}
\frac1{\epsilon^2} \sum_m D_m (\epsilon\pm 1)^{2m},
\end{align}
with coefficients $D_m$,
where $\pm$ corresponds to the axial and vector cases, respectively, and we take into account that the factors $(\epsilon\pm 1)$
can only appear in even powers since the residues of the double-poles have to coincide.
Thus, the difference between the axial-vector and vector operators for a given diagram takes the form
\begin{align}
\frac1{\epsilon}\sum_m 4m D_m+O(\epsilon^0)\,.
\end{align}
The coefficients $D_m$ are sufficiently easy to calculate. We have checked that the difference of the two-loop evolution
kernels~\eqref{HAVdiff} calculated in this way diagrammatically coincides with the result in Eq.~\eqref{deltaHAV}.

The difference of two-loop conformal anomalies~\eqref{Sfull} for axial-vector and vector operators, $\Delta^{(2)}_{A-V} = \Delta^{(2)}_{A}-
\Delta^{(2)}_{V}$ can be written as a sum of two terms
\begin{align}
  \label{widehatDelta}
  \Delta^{(2)}_{A-V}  = \frac14 [\mathbb H_{A-V}^{(2)},z_1+z_2] +  \widehat {\Delta}^{(2)}_{A-V},
\end{align}
where the operator $\widehat {\Delta}^{(2)}_{A-V}$ is defined as
\begin{align}\label{Delta+2}
[\widehat{\Delta}^{(2)}_{A-V } f](\boldsymbol z)  & = \int_0^1 d\alpha\int_0^{\bar\alpha}d\beta
\left[\omega^{(2)}(\alpha)- \omega^{(2)}(\beta) \right]\
 f(z_{12}^\alpha,z_{21}^\beta) ,
\end{align}
with the kernel
\begin{align}
     \omega^{(2)}(\alpha)& =8C_F^2\Big[\frac32\alpha
    +\ln \alpha\left(\frac1 {\bar\alpha}-\bar\alpha\right)
    +\bar\alpha\ln\bar\alpha\Big].
\end{align}
Contributions with the color factors $\beta_0 C_F$ and $C_F C_A$ cancel in the sum of all diagrams.

%
\section{Three-loop evolution kernel}\label{sect:threeH}
%

The canonically non-invariant part of the evolution kernel is completely determined by the commutation relations in
Eq.~\eqref{SHcomm}. The analysis of these equations becomes much simpler after making a similarity transformation of the
operators at the intermediate step~\cite{Braun:2017cih},
\begin{align}
\mathrm S^{\p}_{\alpha} &= \mathrm U^{-1}_{\p}\mathbf S^{\p}_{\alpha}\, \mathrm U_{\p},
&&
\mathbb H_{\p}= \mathrm U^{-1}_{\p}\mathbf H_{\p}\,\mathrm U_{\p},
\end{align}
where the rotation matrix $\mathrm U_{\p}$ is chosen in such a way that
the new (boldfaced) symmetry generators do not include the conformal anomaly term:
\begin{align}
\mathbf S^{\p}_{-} &=S_-,
\notag\\
\mathbf S^{\p}_{0} & = S_0+\left(-\epsilon+\frac12\mathbf H_{\p}\right),
\notag\\
\mathbf S^{\p}_{+} & = S_++(z_1+z_2)\left(-\epsilon+\frac12\mathbf H_{\p}\right).
\end{align}
The rotation matrix can be written in the form~\cite{Braun:2017cih}
\begin{align}
\label{rotation1}
\mathrm U_{\p}=\exp\Big\{a\mathbb X^{(1)}_{\p}+a^2\mathbb X_{\p}^{(2)}+O(a^3)\Big\}.
\end{align}
The one-loop $\mathbb X$-kernels for the vector and axial-vector operators are equal to each other,
\begin{align}
{}[\mathbb X^{(1)}_{\p}f](\boldsymbol z) &= 2C_F\!\int_0^1\!d\alpha \frac{\ln\alpha}{\alpha}\Big[2f(\boldsymbol z)
 -
f(z_{12}^\alpha,z_2)-f(z_1,z_{21}^\alpha)
\Big],
\end{align}
and obey the equation $[S_+,\mathbb X^{(1)}_{\p}]=z_{12}\Delta^{(1)}_{\p}$.
The expression for the two-loop kernel for the vector case, $\mathbb X_V^{(2)}$, can be found
in~\cite{Braun:2017cih}. The difference
\begin{align}
\mathbb X^{(2)}_{A-V}=\mathbb X^{(2)}_{A}-\mathbb X^{(2)}_{V}
\end{align}
is defined as a solution of the equation~\footnote{
Note that equations of the type $[S_+, {\mathcal O}](\boldsymbol z) = {\mathcal G}(\boldsymbol z)$ are nothing but first-order
inhomogeneous differential equations,
whose solutions are defined up to an arbitrary invariant operator, $[S_+,{\mathcal O}^{\text{inv}}]=0$.
Clearly, this ambiguity only results in a redefinition of $\mathrm H^{(3),\mathrm{inv}}_{A-V}$ in Eq.~\eqref{Hseparation}.}
\begin{align}\label{SX2comm}
[S_+,\mathbb X^{(2)}_{A-V}]=\frac14[\mathbb H_{A-V}^{(2)},z_1+z_2]+z_{12}\widehat\Delta^{(2)}_{A-V}.
\end{align}
The solution can be written as a sum of two terms
\begin{align}
\mathbb X^{(2)}_{A-V}=\frac14 \mathbb T_{ A-V}^{(2)} + \Delta\mathbb X^{(2)}_{A-V},
\end{align}
corresponding to the two contributions on the r.h.s.~of Eq.~\eqref{SX2comm},
respectively.
We find
\begin{align}\label{TAV2}
{}[\mathbb T_{A-V}^{(2)}f](\boldsymbol z)=16C_F\beta_0\!\int_0^1d\alpha\!\int_0^{\bar\alpha}\!d\beta\ln(1-\alpha-\beta)
f(z_{12}^\alpha,z_{21}^\beta)
\end{align}
and
\begin{align}
{}[\Delta\mathbb X^{(2)}_{A-V}f](\boldsymbol z) &=\int_0^1 d\alpha \int_0^{\bar\alpha}d\beta\, \big[ \chi(\alpha)+\chi(\beta)\big]
f(z_{12}^\alpha,z_{21}^\beta),
\end{align}
with the weight function
\begin{align}
\chi(\alpha) &=8 C_F^2 \Big[ -\frac12\ln\bar\alpha
 +\frac{\alpha}{\bar\alpha}  \ln\alpha + \Li_2(\bar\alpha) - \Li_2(\alpha)
\Big].
\end{align}
Using these results  for the rotation matrix \eqref{rotation1} and the expression
given in Ref.~\cite[Eq.~(3.12)]{Braun:2017cih} for the three-loop kernel $\mathbb H^{(3)}_V$
we can restore the $O(a^3)$ contribution to the
difference $\mathbb H_{A-V}$, Eq.~\eqref{HAVdiff}. The result reads
\begin{align}
\label{H3A-V}
\mathbb H^{(3)}_{A-V} & = \mathrm H^{(3),\mathrm{inv}}_{A-V} + \frac12 \mathbb T_V^{(1)} \mathbb H^{(2)}_{A-V} +
\mathbb T_{A-V}^{(2)}\left(\beta_0+\frac12 \mathbb H_V^{(1)}\right)
\notag\\
&\quad
+ [\mathbb H_V^{(1)},\mathbb X_{A-V}^{(2)}] + [\mathbb H^{(2)}_{A-V}, \mathbb X_V^{(1)}].
\end{align}
The expressions for   $\mathbb T_{A-V}^{(2)}$, $\mathbb H^{(2)}_{A-V}$, $\mathbb H_V^{(1)}$, $\mathbb X_V^{(1)}$,  $\mathbb X_{A-V}^{(2)}$
are given above and in Ref.~\cite{Braun:2017cih} and
\begin{align}
[\mathbb T_V^{(1)}f](\boldsymbol z)  &= -4C_F\biggl\{
\int_0^1d\!\alpha\frac{\bar\alpha\ln\bar\alpha}{\alpha}\!\Big[
f(z_{12}^\alpha,z_2)+f(z_1,z_{21}^\alpha)\Big]
\notag\\&\quad
+\int_0^1\!d\alpha\int_0^{\bar\alpha}\!d\beta\,\ln(1\!-\!\alpha\!-\!\beta)
f(z_{12}^\alpha,z_{21}^\beta)\biggr\}
\end{align}
is defined as a solution of the equation
\begin{align}
[S_+,\mathbb T_V^{(1)}] & = [\mathbb H^{(1)}_V,z_1+z_2].
\end{align}
The last missing element in Eq.~\eqref{H3A-V} is the invariant kernel $\mathbb H_{A-V}^{(3)}$.
It is completely determined by its eigenvalues
\begin{align}
\mathrm H^{(3),\mathrm{inv}}_{A-V}(z_1-z_2)^{N-1}=\gamma_{A-V}^{(3),\text{inv}}(N)\,(z_1-z_2)^{N-1},
\end{align}
which can be found as
\begin{align}
\gamma_{A-V}^{(3),\text{inv}}(N)=\gamma_{A-V}^{(3)}(N)-\gamma_{A-V}^{(3),\text{ninv}}(N).
\end{align}
Here $\gamma_{A-V}^{(3)}(N)$ is the difference of the three-loop anomalous dimensions of vector and axial-vector operators
(i.e., the eigenvalues of the kernel $\mathbb H^{(3)}_{A-V}$) and $\gamma_{A-V}^{(3),\text{ninv}}(N)$  are the
eigenvalues of the non-invariant operators on the r.h.s.~of Eq.~\eqref{H3A-V}.  One easily  finds (note that the commutator terms do not
contribute to $\gamma^{(3),\text{ninv}}_{A-V} (N)$)
\begin{align}
\gamma^{(3),\text{ninv}}_{A-V} (N) & =
8\beta_0 C_F \frac{d}{d N}\left(\frac{\gamma_V^{(1)}(N)+2\beta_0}{N(N+1)}\right),
\end{align}
where
\begin{align}
\gamma_V^{(1)}(N)=4 C_F\left(S_1(N+1)+S_1(N-1)-\frac32\right)
\end{align}
is the eigenvalue of the  one-loop kernel $\mathbb H_V^{(1)}$
with harmonic sums $S_{\vec m}(N)$, cf.~Ref.~\cite{Vermaseren:1998uu}.
The difference in the three-loop anomalous dimensions, $\gamma_{A-V}^{(3)}(N)$, was calculated
in Ref.~\cite{Moch:2014sna}. The result reads
\begin{align}
\label{gamma3noninv}
\gamma_{A-V}^{(3)}(N)=-2\int_0^1 dx\, x^{N-1} P^{(3)}_{A-V}(x)\,,
\end{align}
with the splitting function written in terms of two functions, $z^{(1)}_{\rm ns}(x)$, $z^{(2)}_{\rm ns}(x)$,
and their convolutions `$\otimes$'
\begin{align}
P^{(3)}_{A-V}(x)=
\beta_1 z^{(1)}_{\rm ns}(x) - \beta_0\left( [z^{(1)}_{\rm ns}\otimes z^{(1)}_{\rm ns}](x)-2 z^{(2)}_{\rm ns}(x)\right),
\end{align}
see~\cite[Eqs.~(A.1), (A.2)]{Moch:2014sna}.
Starting from these expressions, we find the eigenvalues of the invariant kernel after some algebra as
\begin{align}
\label{gamma3inv}
\gamma_{A-V}^{(3),\text{inv}}(N) &
=  \frac{16 C_F \beta_1}
{N(N+1)}
+16
C_F\beta_0\biggl\{ \frac{5}3  \beta_0 \frac1{N(N+1)}
\notag\\
&\quad
+C_A\biggl(\frac{(-1)^{N-1} 2 (2S_{-2}(N)+\zeta_2)}{N(N+1)}-\frac{2\zeta_2}{N(N+1)}
\notag\\
&\quad
+\frac{16}3\frac1{N(N+1)}
+\frac6{N^2(N+1)^2}+\frac4{N^3(N+1)^3}\biggr)
\notag\\
&\quad
-4 C_F\biggl(\frac{(-1)^{N-1} (2S_{-2}(N)+\zeta_2)}{N(N+1)}
+\frac1{N(N+1)}
\notag\\
&\quad
+\frac{13}4\frac1{N^2(N+1)^2}
+\frac2{N^3(N+1)^3}\biggr)
\biggr\}.
\end{align}
It can easily be verified that this expression possesses the so-called reciprocity property
\cite{Dokshitzer:2005bf,Basso:2006nk,Alday:2015eya,Alday:2015ewa}:  its asymptotic expansion at large $N$ is invariant under the
substitution $N\to -N-1$.

The last step is to restore the invariant operator $\mathbb H_{A-V}^{(3),\text{inv}}$ from its spectrum.
Any invariant operator  can be written in the form
\begin{align}
[\mathbb H^{\text{inv}}f](z_1,z_2) &=\int_0^1d\alpha\int_0^{\bar\alpha} d\beta\, h(\tau) f (z_{12}^\alpha,z_{21}^\beta),
\end{align}
where $\tau=\alpha\beta/\bar\alpha\bar\beta$ is the so-called conformal ratio.
The eigenvalues are given by moments of the function $h(\tau)$
\begin{align}
\gamma^{\text{inv}}(N)=\int_0^1d\alpha\int_0^{\bar\alpha} d\beta\, h(\tau)(1-\alpha-\beta)^{N-1},
\end{align}
so that $h(\tau)$ can be restored uniquely by a Mellin transform. Since the expressions for $\gamma_{A-V}^{(3),\text{inv}}(N)$ are
relatively simple, this calculation is rather straightforward. We find
\begin{align}
h_{A-V}^{(3),\text{inv}}(\tau) &= 16C_F\biggl\{\beta_1 +  \beta_0\biggl(\frac53 \beta_0
+4C_F\left[\frac 1 4\ln\bar\tau-\zeta_2+\frac53\right]
\notag\\
&\quad
+\frac2{N_c}\left[\Li_2(\tau) -\zeta_2 +\ln^2\bar\tau-\ln\bar\tau + \frac{8}3
\right]
\biggr)\!\biggr\}.
\end{align}
With this last piece, the difference in the three-loop evolution kernel $\mathbb H_{A-V}^{(3)}$ and, therefore, also
in the axial-vector kernel $\mathbb H_{A}^{(3)}$ itself, is completely fixed.

%
\section{Matching}\label{sect:matching}
%

Vector and axial-vector operators in the $d$-dimensional theory at the critical point have different scaling dimensions, i.e., different
evolution equations in the chosen scheme for $\gamma_5$. In four dimensions, this difference is avoidable: There exist four-dimensional
regularization schemes in which the (flavor-nonsinglet) vector and axial-vector operators satisfy the same evolution equation. Hence it is
possible to define a new operator ${\mathcal O}_A(\boldsymbol z) \mapsto {\mathcal O}_5(\boldsymbol z)$ by the transformation (finite
renormalization)
\begin{align}
{\mathcal O}_5(\boldsymbol z)={\mathcal U}  {\mathcal O}_A(\boldsymbol z)
\label{match}
\end{align}
such that ${\mathcal O}_5(\boldsymbol z)$ obeys the same evolution equation as the vector operator ${\mathcal O}_V(\boldsymbol z)$ in
Eq.~\eqref{vector}, which is convenient for applications and conventionally defines the $\MS$-scheme. In this Section we derive the
explicit expression for the matching kernel~${\mathcal U}$.

Inserting~Eq.~\eqref{match} in  Eq.~\eqref{VARGE} it is easy to see that
the operator ${\mathcal O}_5(\boldsymbol z)$ satisfies the RG equation (in $d=4$) with the
evolution kernel ${\mathbb H}_5$ which is related to $\mathbb H_A$ as follows
\begin{align}
{\mathcal U}^{-1}(a){\mathbb H}_5(a) {\mathcal U}(a)+{\mathcal U}^{-1}(a)\beta(a)\partial_a {\mathcal U}(a)= \mathbb H_{A}(a).
\end{align}
Requiring that ${\mathbb H}_5(a)=\mathbb H_V(a)$ and making an ansatz for the ${\mathcal U}$-kernel in the form
\begin{align}
 {\mathcal U}(a) & = {\text P} \exp\left\{\int_{0}^{a}\frac{d s}{\beta(s)} V(s)\right\}
\\
&=
1 - \frac a{2\beta_0} V_1
  -\frac{a^2}{4\beta_0}\left(V_2-\frac1{2\beta_0} V_1^2 -\frac{\beta_1}{\beta_0} V_1\right)+O(a^3),
 \notag
\end{align}
where $V(s)= s^2 V_1+ s^3 V_2+O(s^4)$, one obtains
\begin{align}\label{VAEQ}
 V(a) &={\mathcal U}(a) \mathbb H_A(a) {\mathcal U}^{-1}(a)-\mathbb H_V(a)
\notag\\
&=a^2\mathbb H_{A-V}^{(2)}+a^3\mathbb H_{A-V}^{(3)}+\frac{a^2}{2\beta_0}[\mathbb H_A^{(1)}, V_1]
\notag\\
&\quad
+\frac{a^3}{2\beta_0}\Big( [\mathbb H_A^{(2)}, V_1] +
\frac12
[\mathbb H_A^{(1)}, V_2]\Big) +O(a^4).
\end{align}
Comparing the leading contributions $O(a^2)$ on  both sides of this relation, one gets
\begin{align}
 V_1& =\mathbb H_{A-V}^{(2)} + \frac{1}{2\beta_0}[\mathbb H_A^{(1)}, V_1]\,.
\end{align}
Note that $\mathbb H_{A-V}^{(2)} \propto \beta_0$, cf.~Eq.~\eqref{deltaHAV}. Assuming that
$V_1$ cannot contain terms $\sim 1/\beta_0$, the only possibility to match the
powers of $\beta_0$ in this relation is to require that
\begin{align}
  V_1& =\mathbb H_{A-V}^{(2)} \qquad\text{and}\qquad [\mathbb H_A^{(1)}, V_1]=0\,.
\end{align}
Note that these two conditions are self-consistent since $\mathbb H_{A-V}^{(2)}$ is an invariant operator;
any two invariant operators commute.

Next, comparing the $O(a^3)$ contributions, one obtains an equation for $V_2$
\begin{align}
\label{V2-1}
V_2 + \frac1{4\beta_0}\Big[V_2,\mathbb{H}_A^{(1)}\Big]=
\mathbb{H}^{(3)}_{A-V}
- \frac1{2\beta_0} \Big[V_1, \mathbb{H}_A^{(2)}\Big].
\end{align}
The kernel $\mathbb{H}^{(3)}_{A-V}$ is given in Eq.~\eqref{H3A-V}, $\mathbb{H}_A^{(1)}=\mathbb{H}_V^{(1)}$, see Eq.~\eqref{Honeloop}
and Ref.~\cite[Eq.~(3.12)]{Braun:2017cih},
\begin{align}
\mathbb{H}_A^{(2)}=\mathbb{H}_A^{(2),\text{inv}} +\mathbb T_1\left(\beta_0+\frac12\mathbb H_V^{(1)} \right)
  + [\mathbb H^{(1)}_V,\mathbb X_V^{(1)}]
\end{align}
with $\mathbb{H}_A^{(2),\text{inv}} = \mathbb{H}_V^{(2),\text{inv}} +  \mathbb{H}^{(2)}_{A-V}$.

Since $V_1$ and $\mathbb H_V^{(1)}$ are invariant operators, the invariant parts of $V_2$ and $\mathbb H_A^{(2)}$ drop out from the
commutators on the l.h.s. and r.h.s. of Eq.~\eqref{V2-1}, respectively. Thus, if we  write
\begin{align}
V_2=\mathbb{H}^{(3),\text{inv}}_{A-V} +  V^\prime_2,
\end{align}
the terms in $\mathbb{H}^{(3),\text{inv}}_{A-V}$ cancel out and one ends up with
the equation for the non-invariant part of the two-loop matching kernel:
\begin{align}\label{VprimeEq}
V^\prime_2 +\frac1{4\beta_0}\Big[V^\prime_2,\mathbb{H}_V^{(1)}\Big]  &=
 \frac12 \mathbb T_V^{(1)} \mathbb H^{(2)}_{A-V} + \mathbb T_{A-V}^{(2)}\Big(\beta_0+\frac12 \mathbb H_V^{(1)}\Big)
\notag\\&\quad
+ \frac14 [\mathbb H_V^{(1)},\mathbb T_{A-V}^{(2)}]
\notag\\&\quad
-\frac1{2\beta_0}[\mathbb H^{(2)}_{A-V},\mathbb T_V^{(1)}]\Big(\beta_0+\frac12\mathbb H_V^{(1)} \Big)
\notag\\&\quad
+ [\mathbb H_V^{(1)},\Delta \mathbb X_{A-V}^{(2)}]+ [\mathbb H^{(2)}_{A-V}, \mathbb X_V^{(1)}]
\notag\\&\quad
-\frac1{2\beta_0}[\mathbb H^{(2)}_{A-V},[\mathbb H^{(1)}_V,\mathbb X_V^{(1)}]].
\end{align}
The expression on the r.h.s. of this equation is a second order polynomial in $\beta_0$ and
we expect the matching kernel ${\mathcal U}$ and, hence, $V(a)$ to be polynomials in $\beta_0$ as well (e.g., since Feynman diagrams are
trivially polynomials in the number of flavors $n_f$).

We write
\begin{align}
V^\prime_2= \beta_0^2 V^\prime_{2,2} + \beta_0 V^\prime_{2,1} + V^\prime_{2,0}
\end{align}
and collect the contributions of different powers of $\beta_0$ separately. One obtains
four equations:
\begin{widetext}
\begin{eqnarray}
  \label{beta+2}
(\beta_0)^{2}: &\,\,\qquad\qquad\qquad\quad\displaystyle
 \beta^2_0 V^\prime_{2,2}  &=\beta_0 \mathbb T^{(2)}_{A-V}\,,
\\
  \label{beta+1}
(\beta_0)^{1}: &\quad\displaystyle
\beta_0 V^\prime_{2,1} + \frac{\beta_0}{4}\Big[V^\prime_{2,2},\mathbb{H}_V^{(1)}\Big] &=
 \frac12 \mathbb T_V^{(1)} \mathbb H^{(2)}_{A-V}
+ \frac12\mathbb T_{A-V}^{(2)}\mathbb H_V^{(1)}
+ \frac14 [\mathbb H_V^{(1)},\mathbb T_{A-V}^{(2)}]
-\frac12[\mathbb H^{(2)}_{A-V},\mathbb T_V^{(1)}]
+ [\mathbb H^{(2)}_{A-V}, \mathbb X_V^{(1)}],
\\
  \label{beta-0}
(\beta_0)^{0}: &\,\,\,\qquad\displaystyle
 V^\prime_{2,0} + \frac1{4}\Big[V^\prime_{2,1},\mathbb{H}_V^{(1)}\Big] &=
[\mathbb H_V^{(1)},\Delta \mathbb X_{A-V}^{(2)}] - \frac1{4\beta_0}[\mathbb H_{A-V}^{(2)}\!,\mathbb T_V^{(1)}]\mathbb{H}_{V}^{(1)}
-\frac1{2\beta_0}[\mathbb H_{A-V}^{(2)}\!,[\mathbb{H}_V^{(1)}\!,\mathbb{X}_V^{(1)}]]\,,
\\
  \label{beta-1}
 (\beta_0)^{-1}: &\qquad\qquad\displaystyle
 \frac1{4\beta_0}\Big[V^\prime_{2,0},\mathbb{H}_V^{(1)}\Big] &= 0
 \,.
\end{eqnarray}
\end{widetext}
From Eqs.~\eqref{beta+2}, \eqref{beta+1} and \eqref{beta-1} we obtain
\begin{align}
\label{Vprime22}
\beta_0V^\prime_{2,2} &= \mathbb T^{(2)}_{A-V}\,,
\\
\label{Vprime21}
\beta_0 V^\prime_{2,1} &=
 \frac12 \mathbb T_V^{(1)} \mathbb H^{(2)}_{A-V}
+ \frac12\mathbb T_{A-V}^{(2)}\mathbb H_V^{(1)}
+ [\mathbb H^{(2)}_{A-V}, \mathbb X_V^{(1)}],
\\
\label{Vprime20}
V^\prime_{2,0} &= 0\,.
\end{align}
The first two expressions follow readily from Eqs.~\eqref{beta+2} and \eqref{beta+1}, respectively.
For the last one, from Eq.~\eqref{beta-1} we conclude that $\Big[V^\prime_{2,0},\mathbb{H}_V^{(1)}\Big]=0$, i.e.
$V^\prime_{2,0}$ is an invariant operator. By virtue of Eq.~\eqref{beta-0}, however, $V^\prime_{2,0}$ is
expressed in terms of commutators that have zero spectrum. Hence $V^\prime_{2,0}$ has zero spectrum
and must vanish, $V^\prime_{2,0}=0$. Collecting all terms we obtain the final result
\begin{align}
\label{V2}
 V_2 &=
\mathbb{H}^{(3),\text{inv}}_{A-V}+ \beta_0 \mathbb T^{(2)}_{A-V}
+ \frac12 \mathbb T_V^{(1)} \mathbb H^{(2)}_{A-V}
+ \frac12\mathbb T_{A-V}^{(2)}\mathbb H_V^{(1)}
\notag\\&\quad
+ [\mathbb H^{(2)}_{A-V}, \mathbb X_V^{(1)}].
\end{align}
Note that Eq.~\eqref{beta-0} was not used to derive the expressions in \eqref{Vprime22}--\eqref{Vprime20} and
provides a highly nontrivial consistency check:
\begin{align}
0 &= \frac1{4}\Big[\mathbb{H}_V^{(1)},V^\prime_{2,1}\Big]
+ [\mathbb H_V^{(1)},\Delta \mathbb X_{A-V}^{(2)}] - \frac1{4\beta_0}[\mathbb H_{A-V}^{(2)},\mathbb T_V^{(1)}]\mathbb{H}_{V}^{(1)}
\notag\\
&\quad
-\frac1{2\beta_0}[\mathbb H_{A-V}^{(2)},[\mathbb{H}_V^{(1)},\mathbb{X}_V^{(1)}]]\,.
\end{align}
Using $V^\prime_{2,1}$ from Eq.~\eqref{Vprime21}, the expression on the r.h.s. of this relation can be brought to
the form  $[\mathbb H_V^{(1)}, \mathbb F]$ where
\begin{align}
\mathbb F &=\Delta \mathbb X_{A-V}^{(2)} -\frac1{8\beta_0}\Big(
2[\mathbb H_{A-V}^{(2)},\mathbb{X}_V^{(1)}]-\mathbb T_V^{(1)} \mathbb H^{(2)}_{A-V}+\mathbb T_{A-V}^{(2)}\mathbb H_V^{(1)}\Big).
\end{align}
%
The equation $[\mathbb H_V^{(1)}, \mathbb F]=0$  implies that $\mathbb F$ is an invariant operator and, therefore, it commutes with the
canonical conformal symmetry generators, $[S_+, \mathbb F]=0$. Since $[S_+,\mathbb X^{(2)}_{A-V}]=\frac14[\mathbb
H_{A-V}^{(2)},z_1+z_2]+z_{12}\widehat\Delta^{(2)}_{A-V}$, cf.~Eq.~\eqref{SX2comm}, it follows from  $[S_+, \mathbb F]=0$  that
\begin{align}
\label{surp}
z_{12}\widehat\Delta^{(2)}_{A-V} & =\frac1{8\beta_0}\Big(
2[\mathbb H_{A-V}^{(2)},z_{12}\Delta_+^{(1)}]-[\mathbb H_V^{(1)},z_1+z_2] \mathbb H^{(2)}_{A-V}
\notag\\
&\quad
+[\mathbb H_{A-V}^{(2)},z_1+z_2]\mathbb H_V^{(1)}\Big).
\end{align}
We have verified that this equation holds true.
Alternatively, Eq.~\eqref{surp} can be used to obtain the conformal anomaly
$\widehat\Delta^{(2)}_{A-V}$
%
avoiding the diagrammatic calculation described in Sect.~\ref{sec:two-loop-anomaly}.

The result for the matching kernel in Eq.~\eqref{V2} can be written as an integral operator in momentum
fraction representation. This is, however, not necessary as the matching can be implemented more efficiently
starting from position-space kernels as explained in Ref.~\cite{Braun:2020yib}. For the applications to LCDAs the
expansion in local operators is more useful. We derive the corresponding expressions in the Appendix.

\section{Summary}\label{sect:summary}

We have calculated the three-loop evolution kernel for flavor-nonsinglet
axial-vector operators in general off-forward kinematics in Larin's scheme.
In QCD applications axial-vector operators are usually defined in such a way that their scale dependence
coincides with that for the vector operators, which can be achieved starting from Larin's scheme and
applying a finite renormalization (matching) to arrive at the conventional choice for the $\MS$-scheme.
Of course, the coefficient functions in the OPE have to be modified accordingly.
We have derived the explicit expression for the matching kernel for the light-ray operators
and also for local operators.

Our method of calculation is based on using the restrictions on off-forward operator mixing
that are due to conformal symmetry of the QCD Lagrangian and can most naturally be
taken into account going over to the Wilson-Fisher critical
point at non-integer $d=4-2\epsilon$ space-time dimensions. In this work the
technique has been generalized to axial-vector operators and we
have derived the two-loop expression for the special conformal anomaly
for axial-vector operators.

The results are relevant for QCD studies of 
hard exclusive reactions involving momentum transfer between the initial and the final states, e.g., DVCS and 
$\gamma^*\gamma\to\pi$ form factor. Concrete applications are beyond the scope of this study.

\section*{Acknowledgments}
This project was supported by Deutsche Forschungsgemeinschaft (DFG) through
the Research Unit FOR 2926, ``Next Generation pQCD for Hadron Structure: Preparing for the EIC'', project number 40824754.
The work of A.M. was supported in addition by the DFG grants MO 1801/4-1, KN 365/13-1 and RSF project No 19-11-0013.

\appendix
\renewcommand{\theequation}{\Alph{section}.\arabic{equation}}


\section{Matching kernel for local operators}\label{sect:appendix}

Light-ray operators are nothing but the generating functions for the renormalized local operators. The results in the local form are
required for several applications, e.g. the calculation of moments of the LCDAs and GPDs using lattice QCD techniques. Our notations in
this Appendix closely follow Sect.~6 in Ref.~\cite{Braun:2017cih}.

Instead of using mixing matrices for the operators with a given number of left and right derivatives,
it proves to be more convenient to go over to the Gegenbauer polynomial basis:
\begin{flalign}
{\mathcal O}^A_{nk}
 &= (\partial_{z_1}+\partial_{z_2})^k C_n^{3/2} \left( \frac{\partial_{z_1}-\partial_{z_2}}{\partial_{z_1}+\partial_{z_2}} \right)
{\mathcal O}_A(z_1,z_2)\bigg|_{z_i = 0}.
&
\label{Onk}
\end{flalign}
Here $k\ge n $ is the total number of derivatives. The rationale for using Gegenbauer polynomials is that any invariant kernel is diagonal
in this basis. Note that the Lorentz spin of the operator with the lowest dimension for given $n$ is $N=n+1$.

The RG equation for the operators ${\mathcal O}^A_{nk}$ has the form
\begin{align}
\left(\mu\frac{\partial}{\partial\mu} + \beta(a) \frac{\partial}{\partial a} \right) {\mathcal O}^A_{nk} =
- \sum_{n'=0}^n \gamma^A_{nn'}\, {\mathcal O}^A_{n'k}\,.
\end{align}
The mixing matrix $\gamma^A_{nn'}$ is triangular and its diagonal elements are equal to the anomalous dimensions
\begin{align}
 \gamma^A_{nn'}  = 0\quad\text{if}\quad n'>n\,, && \gamma^A_{nn} = \gamma_A(n+1)\,.
\end{align}
The light-ray operator can be expanded over local operators as follows
\begin{align} \label{localExpand}
  {\mathcal O}_A(z_1,z_2 ) = \sum _{n=0}^\infty\sum_{k=n}^{\infty} \Phi_{nk}(z_1,z_2) {\mathcal O}^A_{nk}.
\end{align}
The coefficients $\Phi_{nk}(z_1,z_2)$ in this expansion are homogeneous polynomials in $z_1,z_2$ of degree $k$
\begin{align}
\Phi_{nk}(z_1,z_2) =  \omega_{nk} (S_+)^{k-n} z_{12}^n.
\end{align}
The normalization factor reads~~\cite{Braun:2017cih}
\begin{align}
 \omega_{nk} =  2 \frac{2 n + 3}{(k-n)!} \frac{\Gamma(n + 2) }{\Gamma(n + k + 4)}.
\label{Phi-nk}
\end{align}
The action of any integral operator $\bf A$ on the light-ray operator ${\mathcal O}_A$ can be translated into  the matrix form:
\begin{align}\label{Anonlocal1}
  [{\bf A} {\mathcal O}_A](z_1,z_2) &= \sum_{nk} [{\bf A}\Phi_{nk}](z_1,z_2) {\mathcal O}^A_{nk}
  \notag\\
  &=
  \sum_{nk} \Phi_{nk}(z_1,z_2)\sum _{n'k'}  A_{nn'}^{kk'}\,{\mathcal O}_{n'k'},
\end{align}
where the matrix $A_{nn'}^{kk'}$ is defined as
\begin{align}
  [{\bf A}\Phi_{nk}](z_1,z_2) = \sum_{n'k'} A_{n'n}^{k'k} \Phi_{n'k'}(z_1,z_2),
\end{align}
see Ref.~\cite[Sect. 6]{Braun:2017cih} for more details.
Here we only note that if the operator $\bf A$ commutes with the canonical generator of scale transformations $S_0$,
then its matrix elements are nonzero only if $k=k'$,
%
%
$A_{nn'}^{kk'}\equiv \delta_{kk'}A_{nn'}(k)$.
%
%
If, in addition, $\bf A$ commutes with $S_-$, then the matrix elements  $A_{nn'}$ do not depend on $k$, $A_{nn'}(k)=A_{nn'}$.

Our goal in this Appendix is to work out this representation for the matching kernel \eqref{VAEQ}, $V(a) \mapsto [V(a)]^{kk'}_{nn'}$.
Since the operator $V(a)$
satisfies the both requirements, the $k$-indices are redundant
\begin{align}
[V(a)]^{kk'}_{nn'}=\delta_{kk'}V_{nn'}(a).
\end{align}
Following~\cite{Belitsky:1998gc} we split $V_{mn}$ in the diagonal $m=n$ and
and non-diagonal $n>m$ parts,
\begin{align}
V_{mn}(a)=\delta_{mn} V^\text{D}_n(a) +  V^{\text{ND}}_{mn}(a).
\end{align}
The diagonal part is given in terms of the difference in the anomalous dimensions,
\begin{align}
V^\text{D}_n(a)
= a\,\gamma_{A-V}^{(2)}(n+1) + a^2\gamma_{A-V}^{(3)}(n+1)+\cdots
\end{align}
The leading contribution is
\begin{align}
\gamma_{A-V}^{(2)}(n+1)=\frac{16C_F\beta_0}{(n+1)(n+2)}\,,
\end{align}
The three-loop result $\gamma_{A-V}^{(3)}(n+1) = \gamma_{A-V}^{(3)}(N)$ is available
from Ref.~\cite{Moch:2014sna} and we have presented it in the form
\begin{align}
 \gamma_{A-V}^{(3)}(n+1)=\gamma_{A-V}^{(3),\text{inv}}(n+1)+\gamma_{A-V}^{(3),\text{ninv}}(n+1)\,,
\end{align}
see Eqs.~\eqref{gamma3noninv} and \eqref{gamma3inv} for the explicit expressions.
The off-diagonal part of the matching matrix
\begin{align}
V^{\text{ND}}_{mn}(a)=a^2 V^{(2),\text{ND}}_{mn}+O(a^3)
\end{align}
is a new result. We obtain
\begin{align*}
V^{(2),\text{ND}}_{mn} &
    =-\frac{1}{a(m,n)}\biggl\{ \left(
\gamma_{A-V}^{(2)}(m+1)-\gamma_{A-V}^{(2)}(n+1)
\right)
\notag\\
&\quad\times \Big(\Big(\beta_0+\frac12 \gamma_V^{(1)}(n+1)\Big) \mathbf b_{mn} +\mathbf w^{(1)}_{mn}\Big)
\notag \\
&\quad +\frac12\left(
\gamma_V^{(1)}(m+1)-\gamma_V^{(1)}(n+1)
\right)\gamma_{A-V}^{(2)}(n+1)\mathbf b_{mn}
\biggr\}.
\end{align*}
Here
\begin{align}
a(m,n)=(m-n)(m+n+3)
\end{align}
and  the matrices $\mathbf b_{mn}$, $\mathbf w^{(1)}_{mn}$ have the form~\cite{Braun:2017cih}
\begin{align}
\mathbf{b}_{mn}
&=
-2(2n+3)\vartheta_{mn} \,,
\\
  {\bf w}_{mn}^{(1)} &= 4 C_F (2n+3)\, {a}(m,n)
  \notag\\
  &\quad \times \left( \frac{A_{mn} - S_1(m+1)}{(n+1)(n+2)}
  + \frac{2A_{mn}}{{a}(m,n)} \right) \vartheta_{mn}\,,
\end{align}
%
\newpage
\vspace*{-10mm}
\noindent
%
where
\begin{align}
  \vartheta_{mn} =
\begin{cases}
1 \text{ if } m - n > 0 \text{ and even} \\ 0 \text { else}
\end{cases}
\end{align}
and
\begin{align}
 A_{mn} &= S_1\Big(\frac{m+n+2}2\Big) -S_1\Big(\frac{m-n-2}2\Big)
 \notag\\
 &\quad +2S_1(m-n-1)-S_1(m+1).
\end{align}
\newpage
\vspace*{-10mm}
\noindent
%
The first few non-diagonal elements ($0\leq m,n\leq 7$)
for $N_c=3$ and $n_f=4$ are equal to
\begin{align*}
V^{2,\text{ND}}=
\begin{pmatrix}
0 & 0 & 0 & 0 & 0 & 0 & 0\\
0 & 0 & 0 & 0 & 0 & 0 & 0\\
\frac{2000}9 & 0 & 0 & 0 & 0 & 0 & 0\\
0 & \frac{2800}{729} & 0 & 0 & 0 & 0 & 0\\
\frac{4640}{27} & 0 & -\frac{1120}{81} & 0 & 0 & 0 & 0\\
0 & \frac{52000}{1701} & 0 & -\frac{12512}{945} & 0 & 0 & 0\\
\frac{29800}{243}&0&\frac{1580}{243}&0&-\frac{4312}{405} &0&0
\end{pmatrix}.
\end{align*}



%

\end{document}